\documentclass[aps,prb,twocolumn]{revtex4}
\usepackage{graphicx}
\usepackage{color}
\usepackage{dcolumn}
\usepackage{amsmath}
\usepackage{amssymb}
\usepackage{subfigure,amsmath,verbatim,moreverb,bm}
\def\be{\begin{equation}}
\def\ee{\end{equation}}
\def\ber{\begin{eqnarray}}
\def\eer{\end{eqnarray}}
\def\bern{\begin{eqnarray*}}
\def\eern{\end{eqnarray*}}

\def\nablabold{\mbox{\boldmath $\nabla$}}
\def\rv{{\bf r}}

\def\jv{{\bf j}}
\def\Gv{{\bf G}}

\def\pv{{\bf p}}
\def\qv{{\bf q}}

\def\uv{{\bf u}}

\def\0v{{\bf 0}}

\def\nn{\nonumber}

\def\pa{\partial}
\def\Gcal{{\cal G}}
\def\Ncal{{\cal N}}
\begin{document}
\title{Anti-adiabatic limit of the exchange-correlation kernels of an inhomogeneous electron gas}
\author {V. U. Nazarov}
\affiliation{Research Center for Applied Sciences, Academia Sinica, Taipei 115, Taiwan}
\affiliation{Department of Physical Chemistry, Far-Eastern National Technical University, Vladivostok, Russia.}
\author{I. V. Tokatly}
\affiliation{Nano-Bio Spectroscopy group and ETSF Scientific Development Centre, 
  Departamento de F\'isica de Materiales, Universidad del Pa\'is Vasco, 
  Centro de F\'isica de Materiales CSIC-UPV/EHU-MPC, E-20018 San Sebasti\'an, Spain}
\affiliation{IKERBASQUE, Basque Foundation for Science, E-48011 Bilbao, Spain}
\affiliation{Moscow Institute of Electronic Technology, Zelenograd, 124498 Russia}
\author{S. Pittalis and G. Vignale}
\affiliation{Department of Physics, University of Missouri-Columbia,
Columbia, Missouri 65211}
\date{\today}
\begin{abstract}
We express the high-frequency (anti-adiabatic) limit of the exchange-correlation kernels of an inhomogeneous electron gas in terms of the following equilibrium properties:  the ground-state density, the kinetic stress tensor, the pair-correlation function, and the ground-state exchange-correlation potential.  Of these quantities, the first three are amenable to exact evaluation by Quantum Monte Carlo methods, while the last can be obtained from the inversion of the Kohn-Sham equation for the ground-state orbitals.  The exact scalar kernel, in this limit,  is found to be of very long range in space, at variance with the kernel that is used in the standard local density approximation.    The anti-adiabatic xc kernels should be useful in calculations of excitation energies by time-dependent DFT in atoms, molecules, and solids, and provides a solid basis for interpolation between the low- and high-frequency limits of the xc kernels.
\hspace*{7.0cm}
\par
\end{abstract}
\maketitle
\section{Introduction}
The exchange-correlation (xc) kernel is a quantity of central importance both in time-dependent density functional theory (TDDFT) and in time-dependent current density functional theory (TDCDFT).   It is formally  defined as the functional derivative of the time-dependent xc potential $V_{xc}$ with respect to time-dependent density in TDDFT, or as the functional derivative of time-dependent xc vector potential $A_{xc,i}$ with respect to time-dependent current-density $j_j$ in TDCDFT.  In practice, this kernel connects the physical density- and current  response functions of the interacting many-body system to those of a fictitious non-interacting system -- the so-called Kohn-Sham system -- which has the exact ground-state density. More precisely, one has
\be
\chi^{-1}=\chi_{s}^{-1}-f_{xc} - V_C(|\rv-\rv'|)
\label{fxcs_def}
\ee
in TDDFT\cite{Gross-85} and
\be
\hat \chi^{-1}=\hat \chi_{s}^{-1}-\hat f_{xc} - \frac{c}{e \omega^2}\pa  \, V_C(|\rv-\rv'|) \, \pa'
\label{fxc_def}
\ee
in TDCDFT.\cite{Vignale-96}  Here $\chi$ and $\hat \chi$ denote the exact density and current response functions (the second being a tensor, as emphasized by the hat), $\chi_{s}$ and $\hat \chi_{s}$ are their Kohn-Sham counterparts, $\omega$ is the frequency, $c$ is the velocity of light,  $e$ is the absolute value of the charge of electron,
and $V_C(r)=e^2/r$  is the Coulomb interaction.

The great importance of the $f_{xc}$ kernels stems from the fact that the Kohn-Sham response functions usually misrepresent both the response and the excitation spectrum of the system.  Poles of $\chi_{s}$, for example, occur at differences of Kohn-Sham eigenvalues, which are known to be, at least conceptually, unrelated to the true excitation energies of the system.  It is the task of $f_{xc}$ to shift the poles of the KS response functions from these unphysical values to the actual physical ones, and this is a very difficult task, particularly when it comes to band gaps and excitons.

Very few exact results are known about $f_{xc}$.  Even in the simplest case of a homogeneous electron gas (in which case $f_{xc}$ and $\hat f_{xc}$ are functions of $|\rv-\rv'|$ and $\hat f_{xc}$ decomposes into independent longitudinal and transversal components)  very little is known.
The static limit of $f_{xc}(q,\omega)$ has been calculated by variational and diffusion Quantum Monte Carlo (QMC)
 and fitted to simple formulas.~\cite{Moroni95, Corradini98} It provides the basis for the so-called  adiabatic approximation, which is based on the assumption that the exchange-correlation potential  relaxes to the ground-state form on a time scale that is much shorter than the time scale of the evolution of the system.  In the opposite limit of high-frequency, the quantity  $\lim_{\omega\to\infty}f_{xc}(\omega)\equiv f^\infty_{xc}$ is known from exact sum rules to be expressible in terms of the density, the ground-state kinetic energy, and the ground-state pair correlation function (or equivalently the static structure factor), all of which can be obtained from QMC calculations.  This limit provides the basis for an ``anti-adiabatic approximation", in which $f_{xc}$ is interpreted as the stiffness of the ground-state wave function with respect to instantaneous virtual deformations.  This point will be expounded more fully in the following.    

The frequency dependence of the homogeneous  kernels has been the object of several parametrizations, beginning with the one by Gross and Kohn in 1985 \cite{Gross-85} (see also Refs. \onlinecite{Dabrowski-86} and \onlinecite{Qian-02}), which interpolate between low- and high-frequency limit, and incorporate as many exact properties and constraints as possible.  Perturbative calculations, which treat electron exchange exactly, have also been carried out. \cite{Sturm00,Kim-02}

Historically, the xc kernels of the homogeneous electron gas have played (and still play) an important role as the basis for local density approximations, whereby the xc kernels of the non-homogeneous systems are approximated in terms of those of a homogeneous system, evaluated at the local density.
However, it has become clear in recent years that some features of the exact kernels cannot be  reproduced by the local density approximation.  For example, it is now known that the exact scalar xc kernel has a singularity at finite frequency and $q \to 0$ (corresponding to long range in space), which plays an important role in the optical spectra of extended system. Reining {\it et al.} \cite{Reining-02}  introduced phenomenologically a singular $f_{xc}$, and using it ``on top" of  bands calculated by the GW method demonstrated considerable improvement in the calculated optical spectrum of silicon.
One way to introduce a singular $f_{xc}$ from first principles is to derive it from a local density approximation for the tensor $\hat f_{xc}$ of TDCDFT.  Indeed, any regular approximation for $\hat f_{xc}$, when translated into an approximation for $f_{xc}$, does produce a singularity of the expected form.\cite{Nazarov-07,Nazarov-08}  The extent to which this may help to resolve the difficulties in the calculation of optical spectra remains unclear.   Further it now appears that not only the singular behavior of $f_{xc}$ at $q=0$, but also the detailed non-local spatial dependence of this kernel plays a role in the calculation of the optical spectra.

In view of these difficulties it seems particularly timely to try and learn as much as possible about the exact properties of $f_{xc}$ in non-uniform systems.  In a recent paper we have taken a first step in this direction by deriving an exact expression for the $f_{xc}$ of a weakly inhomogeneous system to second-order in the strength of the external potential.\cite{Nazarov-09}   This expression involves the kernels of the homogeneous electron gas (which, as discussed above,  are still poorly described), but is otherwise fully nonlocal.  Unfortunately, the perturbative approach restricts us to metals, and seems to exclude a priori the interesting case of semiconductors.

In this paper we calculate the exchange-correlation kernels for arbitrary non-homogeneous systems in the anti-adiabatic limit.  As remarked above, the anti-adiabatic limit of the {\it homogeneous} exchange-correlation kernel is well known  (a state-of-the-art formula for it is provided in the appendix).  Here we extend that knowledge to the considerably more complex case of a non-homogeneous system.  In spite of the additional complication, we are still able to express the result in terms of ground-state density, kinetic stress tensor, pair correlation function and exact exchange-correlation potential.  All these quantities are well within the reach of modern numerical algorithms, and therefore our results open the way to an accurate study of $f_{xc}$  in the anti-adiabatic limit.

The high-frequency response of quantum many-body systems has recently come under intense scrutiny in the context of the continuum mechanics formulation of quantum many-body theory.\cite{Tao-09}  In this theory one attempts to describe the dynamics in terms of a displacement field $\uv(\rv,t)$,  describing the displacement of an infinitesimal volume element from its ``equilibrium position" in the ground-state  (in this approach the time-evolution is described as a geometric deformation of the ground-state).   It turns out that the simplest approximation that leads to an eigenvalue problem for the excitation energies is in fact equivalent to a high-frequency expansion for the current response function.  We get
\bern
m \ddot\uv(\rv,t)&=&  - \nablabold V_1(\rv,t)\nn\\
& -& \frac{1}{n_0(\rv)}\int d\rv' {\bf M}(\rv,\rv') \cdot \uv(\rv',t)\,,
\eern
where $n_0(\rv)$ is the ground-state density, and the tensor ${\bf M}(\rv,\rv')$ is the first moment of the current excitation spectrum [see Eq.~(\ref{Main_comm})].
This quantity can be expressed in terms of ground-state properties, i.e., it is the second variational derivative of the ground-state energy with respect to the instantaneous displacement field $\uv(\rv)$, evaluated at $\uv = 0$.  We will see that ${\bf M}(\rv,\rv')$ also plays  a crucial role in the construction of $f_{xc}^\infty$.

This paper is organized as follows.  In Section \ref{General} we present a general formulation for the anti-adiabatic tensor and scalar xc kernels  in terms of the first spectral moment of the current density fluctuation spectrum.  In Section \ref{TSxc} we provide explicit expressions for these kernels in finite or infinite systems in real space.  In Section~\ref{ContMech} we make  connection with the physically instructive continuum mechanics formulation, and discuss the interpretation  of $f_{xc}$ in terms of  elastic constant (stiffnesses) of the electronic system.    In Section \ref{Periodic} we present the expressions for the tensor and scalar xc kernels for an infinite periodic system in momentum space.       Working in momentum space allows us to focus most clearly on the long-rangedness of the kernels in the $q\to 0$ limit:  Section \ref{qzero} summarizes our results for the singular parts of the kernel.     Finally, Appendix~\ref{D_Mxc} presents a detailed derivation of the all-important double commutator $M$,  and Appendix~\ref{Stefano}  presents results for the longitudinal and transversal component of the anti-adiabatic $\hat f_{xc}$ in the homogeneous electron gas, obtained by using  state-of-the-art values of the static structure factor and correlation energy.

\section{General formulation}
\label{General}

The  current response tensor can be written as
\ber
&&\chi_{\mu\nu}(\rv,\rv',\omega)=
\frac{e n_0(\rv)  \delta(\rv-\rv')}{c m} \delta_{\mu\nu}\nn\\&&+ e \sum\limits_{\alpha,\beta} [F(\alpha)-F(\beta)] 
\frac{\langle \alpha | j_\mu(\rv)|\beta\rangle \langle \beta | j_\nu(\rv')|\alpha\rangle }{\omega-E_\beta+E_\alpha+i \eta},
\label{chi_basic}
\eer
where $|\alpha\rangle$ and $E_\alpha$ are the many-body wave functions and eigenenergies, respectively, $n_0(\rv)$ is the ground-state electron density,
$\jv(\rv)=\sum_{n} [\pv_n \delta(\rv-\rv_n)+\delta(\rv-\rv_n) \pv_n]/(2 m)$
is the paramagnetic current-density operator, $\pv_n$ is  the momentum operator of the $n$-th electron, $F(\alpha)$ are the diagonal elements of the 
equilibrium (or ground-state at $T=0$) density matrix,
$\omega$ is the frequency, $c$ is the velocity of light, $e$ and $m$ are the absolute value of the charge and mass of electron, respectively.
Expansion of the right-hand side of Eq.~(\ref{chi_basic}) in $1/\omega$ gives to the second order
\ber
\chi_{\mu\nu}(\rv,\rv',\omega)\stackrel{\omega \to \infty}{\approx} \frac{e}{c} \left[\frac{n_0(\rv)}{m} \delta(\rv-\rv') \delta_{\mu\nu}
+\frac{M_{\mu\nu}(\rv,\rv') }{m^2 \omega^2} \right],
\label{chi-M}
\eer
where
\ber
M_{\mu\nu}(\rv,\rv')=m^2 \langle [j_\mu(\rv),[H_0,j_\nu(\rv')]]\rangle
\label{Main_comm}
\eer
is the first moment of the imaginary part of $\chi_{\mu\nu}$,
the angle brackets denoting the ground-state or equilibrium average at zero or finite temperature, respectively,
\bern
H_0=\sum\limits_n \left[\frac{ p_n^2}{2 m}   + V_0(\rv_n)\right] + \frac{1}{2} \sum\limits_{n\ne m} V_C(|\rv_n-\rv_m|)
\eern
is the Hamiltonian of the unperturbed system with $V_0(\rv)$  being the external  potential. 
Inversion of Eq.~(\ref{chi-M}) yields
\bern
\chi^{-1}_{\mu\nu}(\rv,\rv',\omega)\stackrel{\omega \to \infty}{\approx} \frac{c}{e} \left[\frac{m \delta(\rv-\rv') }{ n_0(\rv) }\delta_{\mu\nu} - 
\frac{M_{\mu\nu}(\rv,\rv')}{\omega^2 n_0(\rv) n_0(\rv')} \right].
\eern

Using the definition of the tensor exchange-correlation kernel of TDCDFT\cite{Vignale-96} of Eq.~(\ref{fxc_def})
we can write
\ber
f_{xc,\mu\nu}(\rv,\rv',\omega) \stackrel{\omega \to \infty}{\approx}
\frac{c  M^{xc}_{\mu\nu}(\rv,\rv')}{e \omega^2 n_0(\rv) n_0(\rv')},
\label{ftM}
\eer
where 
\ber
 M^{xc}_{\mu\nu}(\rv,\rv') &=&  M_{\mu\nu}(\rv,\rv') - M^s_{\mu\nu}(\rv,\rv') \nn\\&-& n_0(\rv) n_0(\rv') \pa_\mu \pa'_\nu V_C(|\rv-\rv'|).
\label{Mxc}
\eer

The connection to the  scalar quantities follows from the equality
\be
\chi(\rv,\rv',\omega)= \frac{c\, \pa_\mu  \pa'_\nu \chi_{\mu\nu}(\rv,\rv',\omega)}{e \, \omega^2}
,
\label{chit}
\ee
where $\chi(\rv,\rv',\omega)$ is the density response function, and the summation over the  repeated 
coordinate indices is implied.
By virtue of Eqs.~(\ref{chi-M}) and (\ref{chit}) we can write
\ber
\chi(\rv,\rv',\omega)&\stackrel{\omega \to \infty}{\approx}& \frac{1}{m \omega^2}  \pa_\mu n_0(\rv)\pa'_\mu \delta(\rv-\rv') \nn\\ &+& \frac{\pa_\mu \pa'_\nu M_{\mu\nu}(\rv,\rv')}{ m^2 \omega^4},
\label{chi-Ms}
\eer
the inversion of which gives
\ber
&&\chi^{-1}(\rv,\rv',\omega)\stackrel{\omega \to \infty}{\approx} m \omega^2 \Gcal(\rv,\rv')\nn\\ &&- \int [\pa''_\mu \Gcal(\rv,\rv'')]   M_{\mu\nu} (\rv'',\rv''') [\pa'''_\nu \Gcal(\rv'',\rv''')]  d\rv'' d\rv''',\nonumber\\
\label{chi-m1}
\eer
 where $\Gcal(\rv,\rv')=[\pa_\mu n_0(\rv) \pa'_\mu \delta(\rv-\rv')]^{-1}$ can be written as the solution of the following equation
\be
-\pa_\mu n_0(\rv) \pa_\mu  \Gcal(\rv,\rv')= \delta(\rv-\rv').
\ee

With the use of the definition of the scalar xc kernel of Eq.~(\ref{fxcs_def})
we have from Eqs.~(\ref{chi-m1}) and (\ref{Mxc})
\be
f^\infty_{xc}(\rv,\rv') \! = \! \! \int \! [\pa''_\mu \Gcal(\rv,\rv'')]   M^{xc}_{\mu\nu} (\rv'' \! ,\! \rv''') [\pa'''_\nu \Gcal(\rv'' \! ,\rv''')]  d\rv'' d\rv'''\,,
\label{fxc_sc}
\ee
where  $f^\infty_{xc}(\rv,\rv') \equiv \lim_{\omega\to\infty}f_{xc}(\rv,\rv',\omega)$.

\section{Tensor and scalar xc kernels in real space}
\label{TSxc}
In Appendix \ref{D_Mxc}, we outline a rather lengthy evaluation of the double commutator in Eq.~(\ref{Main_comm}) which leads to the following 
final expressions
\ber
&&M^{xc}_{\mu\nu}(\rv,\rv')= T^{xc}_{\alpha\alpha\mu\nu}(\rv,\rv')+T^{xc}_{\nu\alpha\alpha\mu}(\rv,\rv')+T^{xc}_{\alpha\mu\alpha\nu}(\rv,\rv')\nn\\
&&+ \delta(\rv-\rv') \left[\int K^{xc}_{\mu\nu}(\rv,\rv'') d\rv'' -n_0(\rv) \pa_\mu \pa_\nu V_{xc}(\rv)\right]\nn\\
&&-  K^{xc}_{\mu\nu}(\rv,\rv')\,,
\label{fxc_fin}
\eer
where
\be
T^{xc}_{\alpha\beta\mu\nu}(\rv,\rv')=\pa_\alpha\pa'_\beta \delta(\rv-\rv') T^{xc}_{\mu\nu}(\rv),
\ee
\ber
T^{xc}_{\mu\nu}(\rv) \! = \! \frac{1}{2 m} \left(\pa_\mu \pa'_\nu \! + \! \pa'_\mu\pa_\nu) [ \rho_1(\rv,\rv') \! - \!  \rho_{1S}(\rv,\rv') ]\right|_{\rv'=\rv}, \ \
\label{Txc}
\eer
$\rho_1(\rv,\rv')$ is one-particle density-matrix, 
\be
 K^{xc}_{\mu\nu}(\rv,\rv') =  n_0(\rv)  n_0(\rv') [g(\rv,\rv')-1]  \pa_\mu \pa_\nu V_C(|\rv-\rv'|),
 \label{Kxc}
\ee
$g(\rv,\rv')$ is the pair correlation function defined as\cite{Giuliani&Vignale}
\be
g(\rv,\rv')=\frac{N(N-1) }{n_0(\rv) n_0(\rv')} \langle \delta(\rv-\rv_1)\delta(\rv'-\rv_2) \rangle,
\label{g}
\ee
where $\rv_1$ and $\rv_2$ are the coordinates of two distinct electrons of the system, and $V_{xc}(\rv)$ is the static xc  potential.

For the scalar $f_{xc}$, the substitution of Eq.~(\ref{fxc_fin}) into  Eq.~(\ref{fxc_sc}) leads to
\ber
&&f^\infty_{xc}(\rv,\rv')= \int 3 [\pa''_\alpha \pa''_\mu  \Gcal(\rv,\rv'')]  
   T^{xc}_{\mu\nu}(\rv'') 
 [\pa''_\alpha \pa''_\nu \Gcal(\rv'',\rv')]  d\rv'' \nn\\
&&+\int [\pa''_\mu  \Gcal(\rv,\rv'')]  
  \Omega^{xc}_{\nu\mu}(\rv'')
 [\pa''_\nu  \Gcal(\rv'',\rv')]  d\rv''\nn\\
&&-\int [\pa''_\mu  \Gcal(\rv,\rv'')]   K^{xc}_{\mu\nu}(\rv'',\rv''') 
 [\pa'''_\nu  \Gcal(\rv''',\rv')]  d\rv'' d\rv''', \ \ \ \ \ \
\label{fxc-fin}
\eer
where an effective ``xc curvature'' $\Omega_{\mu\nu}^{\rm{xc}}(\rv)$ is defined as
\ber
\label{Omega}
\Omega^{xc}_{\nu\mu}(\rv)=
\int   K^{xc}_{\mu\nu}(\rv,\rv')  d\rv' -n_0(\rv)   \pa_\mu \pa_\nu V_{xc}(\rv). 
\eer

\section{Formulation in terms of continuum mechanics}
\label{ContMech}
In spite of their complicated appearance, the physical significance of formulas (\ref{fxc_fin})- (\ref{Omega})  is quite transparent.  In fact, the formulas could be derived more quickly starting from the representation of $M^{xc}_{\mu\nu}(\rv,\rv')$ as the second functional derivative of the exchange-correlation energy functional $E_{xc}[\uv]$ with respect to virtual displacements, $u_\mu(\rv)$, $u_\nu(\rv')$.   The explicit expression for $E_{xc}[\uv]$ can be easily obtained from the formulas given in Refs.~\onlinecite{Tao-09} and ~\onlinecite{Gao-10} for the full energy functional $E[\uv]$ (kinetic plus Coulomb interaction plus external potential energy) by subtracting the non-interacting kinetic energy functional, the  Hartree energy functional, and the Kohn-Sham potential energy. 
The main idea behind this ``continuum mechanics" approach is to introduce a deformed ground-state wave function\footnote{We focus on the zero-temperature case for simplicity, but it is evident that everything we say can be easily generalized to an equilibrium ensemble of states at finite temperature.}
 \be
 \psi_0[\uv] (\rv_1,...\rv_N)\equiv \psi_0(\rv_1-\uv(\rv_1),...,\rv_N-\uv(\rv_N)) \prod_{i=1}^N g^{-1/4}(\rv_i)
 \ee
where the factors $g^{-1/4}$ ensure the correct normalization of the deformed state, $g^{1/2}$ being the Jacobian of the non-volume-preserving transformation of coordinates $\rv\to\rv-\uv(\rv)$, i.~e. the determinant of  the matrix $\delta_{\mu\nu}-\partial_\mu u_\nu$. Then we have
 \be
 M_{\mu\nu}(\rv,\rv')=\left.\frac{\delta^2 E[u]}{\delta u_\mu(\rv)\delta u_\nu(\rv')}\right\vert_{\uv=0}\,,
 \ee
where $E[\uv]$ is the expectation value of the Hamiltonian in the deformed ground-state $\psi_0[\uv]$.  Similarly, we have
 \be
 M^{xc}_{\mu\nu}(\rv,\rv')=\left.\frac{\delta^2 E_{\rm{xc}}[\uv] }{\delta u_\mu(\rv)\delta u_\nu(\rv')}\right\vert_{\uv=0}\,,
 \ee
where,  as noted above,  $E_{\rm{xc}}[\uv]$ is defined as the difference $E[\uv]-E_H[\uv]-T_s[\uv] -V_s[\uv]$, where $E_H[\uv]$ is the Hartree energy, $T_s[\uv]$ is the kinetic energy of the deformed Kohn-Sham ground-state (i.e., the non-interacting ground state that has the same density as the true ground-state),  and $V_s[\uv]$ is the Kohn-Sham potential energy in the deformed ground state.

The above expressions are very helpful in understanding the physical significance of the terms that appear in the evaluation of the double commutator (\ref{Main_comm}).  To second order in $\uv$ the xc energy functional has the form\cite{Tao-09,Gao-10}
\begin{widetext}
\ber
\label{IlyaEXC2}
E_{\rm{xc}}[\uv] 
&=&\frac{1}{2}\int d\rv\left\{-u_{\mu}(\rv)u_{\nu}(\rv)n_0(\rv)\partial_{\mu}\partial_{\nu}V_{\rm{xc}}(\rv)
+T_{\mu\nu}^{\rm{xc}}(\rv)[4u_{\mu\alpha}(\rv)u_{\nu\alpha}(\rv) - (\partial_{\mu}u_{\alpha}(\rv))(\partial_{\nu}u_{\alpha}(\rv))] \right\}\nn\\
&+& \frac{1}{4}\int d\rv d\rv'[u_{\mu}(\rv)-u_{\mu}(\rv')]K_{\mu\nu}^{\rm{xc}}(\rv,\rv')[u_{\nu}(\rv)-u_{\nu}(\rv')]\,,
\eer
\end{widetext}
where $T_{\mu\nu}^{\rm{xc}}(\rv)$ and $K_{\mu\nu}^{\rm{xc}}(\rv,\rv')$ are defined by Eqs.~(\ref{Txc}) and (\ref{Kxc}) respectively. 
The second term in (\ref{IlyaEXC2}) is a typical energy of an elastic medium, which is quadratic in the deformation gradients (strains).\cite{Landau-VII} The corresponding elastic moduli are determined by the kinetic stress tensor $T_{\mu\nu}$. The third term is an additional nonlocal contribution which should be present if the continuum is charged and the displacement produces a local polarization. Hence the third term corresponds to the xc part of the dipole-dipole interaction energy of an inhomogeneously polarized continuum. Apparently the first two terms vanish in the case of a homogeneous (rigid) displacement since it does not cause deformations and thus does not cost any energy. For the rigid displacement only the first term in (\ref{IlyaEXC2}) survives. This term ensures the harmonic potential theorem. It cancels the xc part of the KS potential, which guaranties that the rigid motion of the whole system is controlled solely by the external potential.

The above expression for $E_{\rm{xc}}$ can be rewritten in terms of the effective curvature $\Omega_{\mu\nu}^{\rm{xc}}(\rv)$, defined in Eq.~(\ref{Omega}), and we get
\ber
\label{IlyaEXC}
&&E_{\rm{xc}}[\uv]=-\frac{1}{2}\int d\rv d\rv' u_{\mu}(\rv)K_{\mu\nu}^{\rm{xc}}(\rv,\rv')u_{\nu}(\rv')\nn\\
&&+\frac{1}{2}\int d\rv\left\{\Omega_{\mu\nu}^{\rm{xc}}u_{\mu}u_{\nu} +T_{\mu\nu}^{\rm{xc}}[4u_{\mu\alpha}u_{\nu\alpha} - (\partial_{\mu}u_{\alpha})(\partial_{\nu}u_{\alpha})]\right\}\,,\nn\\ 
\eer
from which the expression~(\ref{fxc_fin}) for $M_{xc}$ can be straightforwardly recovered by isolating the coefficient of $u_\mu(\rv) u_\nu(\rv')$ (this requires some integrations by parts).

\section{Application to the periodic case}
\label{Periodic}
In the case of a periodic system, the response functions and xc kernels become infinite matrices 
indexed by reciprocal lattice vectors.
The Fourier transform of Eq.~(\ref{ftM}) gives
\begin{eqnarray}
&& f^{xc}_{\Gv\Gv',\mu\nu}(\qv,\omega) \stackrel{\omega \to \infty}{\approx} \frac{c}{e \omega^2} \times \cr\cr
&&  n_0^{-1}(\Gv \! - \! \Gv'')  M^{xc}_{\Gv''\Gv''',\mu\nu}( \qv) n_0^{-1}(\Gv''' \! \! - \! \Gv'), \ \ \ \ \ \
\label{f_xct}
\end{eqnarray}
where $n_0^{-1}(\Gv)$ is the Fourier transform of $1/n_0(\rv)$ and we imply summation over the repeated reciprocal vectors. By Eq.~(\ref{fxc_fin}), we can write in the reciprocal space 
\begin{widetext}
\ber
&& M^{xc}_{\Gv\Gv',\mu\nu}(\qv)=(\Gv  +  \qv)  \cdot  (\Gv'  +  \qv)
 T^{xc}_{\mu\nu}(\Gv  -  \Gv')
  +   (G'_\alpha  +  q_\alpha)  T^{xc}_{\alpha\mu}(\Gv  -  \Gv') (G_\nu  +  q_\nu)  +  \cr\cr
&&(G'_\mu  +  q_\mu) (G_\alpha  +  q_\alpha)T^{xc}_{\alpha\nu}(\Gv  -  \Gv')-\left[K^{xc}_{\Gv\Gv',\mu\nu} (\qv)-K^{xc}_{\Gv\0v,\mu\nu} (\0v) \delta_{\0v\Gv'}\right]\cr\cr 
&&+ (G'_\mu  -G''_\mu)(G'_\nu  -  G''_\nu) n_0(\Gv  -  \Gv'') V_{xc}(\Gv''   -   \Gv'),
  \label{Mxcper}
\eer
where $T^{xc}_{\mu\nu}(\Gv)$ is the Fourier transform of $T^{xc}_{\mu\nu}(\rv)$ and $K^{xc}_{\Gv\Gv',\mu\nu}(\qv)$
is the double Fourier transform of $ K^{xc}_{\mu\nu}(\rv,\rv') $. 

We can write
\be
f^{xc,\infty}_{\Gv\Gv'}(\qv)=\Gcal_{\Gv\Gv''}(\qv)  M^{xc}_{\Gv''\Gv'''}(\qv) \Gcal_{\Gv'''\Gv'}(\qv),
\label{fxc1}
\ee
where
\bern
M^{xc}_{\Gv\Gv'}(\qv)=(G_\mu+q_\mu)  M^{xc}_{\Gv\Gv',\mu\nu}(\qv) (G'_\nu+q_\nu)= (G_\mu+q_\mu) \times \cr\cr
 \left\{
3 (\Gv  +  \qv)  \cdot  (\Gv'  +  \qv)
 T^{xc}_{\mu\nu}(\Gv  -  \Gv')
 -\left[K^{xc}_{\Gv\Gv',\mu\nu} (\qv)-K^{xc}_{\Gv\0v,\mu\nu} (\0v) \delta_{\0v\Gv'}\right] \right. \cr\cr
\left.
+ (G'_\mu  -G''_\mu)(G'_\nu  -  G''_\nu) n_0(\Gv  -  \Gv'') V_{xc}(\Gv''   -   \Gv')
\right\}
(G'_\nu+q_\nu) .
\eern
\end{widetext}

\section{Long-wavelength limit}
\label{qzero}
Because of the applications in optics and because of the singularities arising in the scalar $f^{xc}$, 
the long-wave limit ($\qv\rightarrow\0v$) requires a separate consideration.

By the structure of Eq.~(\ref{Mxcper}),
$M^{xc}_{\Gv\Gv',\mu\nu}(\qv)$  have finite limit as $\qv\rightarrow\0v$.
By Eq.~(\ref{f_xct}), the same is true for $f^{xc}_{\Gv\Gv',\mu\nu}(\qv)$.
We note that  in the long-wave limit the ``head'' and ``wing'' elements of $M^{xc}_{\Gv\Gv',\mu\nu}$ reduce to 
\bern
 M^{xc}_{\Gv\0v,\mu\nu}= M^{xc}_{\0v\Gv,\nu\mu}= G''_\mu  G''_\nu n_0(\Gv  -  \Gv'') V_{xc}(\Gv''  ),
\eern
i.e., they are expressed in terms of the ground-state density and xc potential only.
Equation (\ref{f_xct}), however, shows that this is not the case with $f^{xc}_{\0v\0v,\mu\nu}$ and $f^{xc}_{\Gv\ne\0v,\0v,\mu\nu}$, for the knowledge of  which 
 $T^{xc}_{\Gv\Gv',\mu\nu}$ and $K^{xc}_{\Gv\Gv',\mu\nu}$ are needed.

For the scalar $f^{xc}$, it can be, however,  noticed that Eq.~(\ref{fxc1})  gives a singular long-wave limit. 
Indeed, since $\Gcal_{\Gv\Gv'}(\qv)=\Ncal^{-1}_{\Gv\Gv'}(\qv)$, where
\be
\Ncal_{\Gv\Gv'}(\qv)=(\Gv+\qv)\cdot(\Gv'+\qv) n_0(\Gv-\Gv')
\ee
and the upper row (left column) of $\Ncal_{\Gv\Gv'}(\qv)$ is zero in the $\qv\rightarrow\0v$ limit, $\Gcal_{\Gv\Gv'}(\qv)$ is singular in this limit.
To isolate the singularity, we introduce a matrix
\bern
\tilde\Ncal_{\Gv\Gv'}(\qv) = \frac{\Ncal_{\Gv\Gv'}(\qv)}{|\Gv+\qv||\Gv'+\qv|}
\eern
the inverse of which is regular at $\qv\rightarrow\0v$.  We can write
\be
f^{xc,\infty}_{\Gv\Gv'}(\qv)=\frac{\tilde\Ncal^{-1}_{\Gv\Gv''}(\qv)  M^{xc}_{\Gv''\Gv'''}(\qv)\tilde\Ncal^{-1}_{\Gv'''\Gv'}(\qv)}{|\Gv+\qv||\Gv''+\qv||\Gv'''+\qv||\Gv'+\qv|}.
\label{fxc2}
\ee

Retaining the leading terms in $1/q$ only, we have in the long-wave limit
\be
f^{xc,\infty}_{\0v\0v}(\qv\rightarrow\0v)=\frac{\alpha}{q^2},
\ee
where
\begin{widetext}
\bern
\alpha=
\left[\tilde\Ncal^{-1}_{\0v\0v}(\hat \qv)\right]^2 \hat q_\mu  M^{xc}_{\0v\0v,\mu\nu} \hat q_\nu +2  \tilde\Ncal^{-1}_{\0v\0v}(\hat \qv) \sum\limits_{\Gv\ne\0v} \! \! \tilde\Ncal^{-1}_{\0v\Gv}(\hat \qv) \hat G_\mu  M^{xc}_{\Gv\0v,\mu\nu} \hat q_\nu + \! \! \!
\sum\limits_{\Gv,\Gv'\ne\0v} \tilde\Ncal^{-1}_{\0v\Gv}(\hat \qv) \hat G_\mu  M^{xc}_{\Gv\Gv',\mu\nu} \hat G'_\nu  \tilde\Ncal^{-1}_{\Gv'\0v}(\hat \qv), 
\eern
\bern
f^{xc,\infty}_{\Gv\ne\0v,\0v}(\qv\rightarrow\0v)&=&f^{xc,\infty}_{\0v,-\Gv}(\qv\rightarrow\0v)\cr\cr
&=&\frac{1}{q G} \left\{\tilde\Ncal^{-1}_{\Gv\0v}(\hat \qv) \hat q_\mu  M^{xc}_{\0v\0v,\mu\nu} \hat q_\nu \tilde\Ncal^{-1}_{\0v\0v}(\hat \qv)+ 
\sum_{\Gv'\ne\0v} \left[\tilde\Ncal^{-1}_{\Gv\Gv'}(\hat \qv)  \tilde\Ncal^{-1}_{\0v\0v}(\hat \qv)+
+\tilde\Ncal^{-1}_{\Gv\0v}(\hat \qv) \tilde\Ncal^{-1}_{\Gv'\0v}(\hat \qv) \right] \hat G'_\mu  M^{xc}_{\Gv'\0v,\mu\nu} \hat q_\nu \right. \nn\\
&+&\left.\sum_{\Gv',\Gv''\ne\0v}\tilde\Ncal^{-1}_{\Gv\Gv'}(\hat \qv) \hat G'_\mu M^{xc}_{\Gv'\Gv'',\mu\nu} \hat G''_\nu \tilde\Ncal^{-1}_{\Gv''\0v}(\hat \qv) \right\}
\eern
\bern
f^{xc,\infty}_{\Gv\ne\0v,\Gv'\ne\0v}(\qv\rightarrow\0v)&=&\frac{1} {GG'} \left\{ \tilde\Ncal^{-1}_{\Gv\0v}(\hat \qv) \hat q_\mu  M^{xc}_{\0v\0v,\mu\nu} \hat q_\nu \tilde\Ncal^{-1}_{\0v\Gv'}(\hat \qv)\right. \nn\\&+&\sum\limits_{\Gv''\ne\0v}\left[\tilde\Ncal^{-1}_{\Gv\Gv''}(\hat  \qv)   \tilde\Ncal^{-1}_{\0v\Gv'}(\hat \qv)+
\tilde\Ncal^{-1}_{\Gv\0v}(\hat \qv) \tilde\Ncal^{-1}_{\Gv''\Gv'}(\hat \qv) \right]\hat G''_\mu M^{xc}_{\Gv''\0v,\mu\nu} \hat q_\nu 
\\
&+&
\left.
\sum\limits_{\Gv'',\Gv'''\ne\0v} \tilde\Ncal^{-1}_{\Gv\Gv''}(\hat \qv)  \hat G''_\mu  M^{xc}_{\Gv''\Gv''',\mu\nu} \hat G'''_\nu \tilde\Ncal^{-1}_{\Gv'''\Gv'}(\hat  \qv)
\right\},\\
\eern
\end{widetext}
$M^{xc}_{\Gv\Gv',\mu\nu}$ are taken at  $\qv=\0v$ and vectors with a hat denote unit vectors in the same direction.

 The singularity in $q$ of the ``head'' and ``wings'' elements of the scalar $f^{xc}$  in the long-wave limit has an  important physical implication:
it is directly relevant to the description of the excitonic effect in semiconductors and insulators by means of TDDFT.\cite{Reining-02,Del_Sole-03}
We point out that this singularity is entirely due to the inhomogeneity,
since it disappears in the homogeneous case when the non-diagonal elements of the matrices are zero (cf. Ref. \onlinecite{Nazarov-09}).

\section{Summary and conclusions}
In the high frequency limit, we have worked out the dynamical exchange-correlation kernel of an inhomogeneous many-body system in terms of the ground-state (or equilibrium at finite temperature) properties of the system. Both the scalar xc kernel $f_{xc}$ and tensor xc kernel $\hat{f}_{xc}$, relevant to the time-dependent density-functional theory and the time-dependent {\em current-density} functional theory, respectively, have been obtained.

The ground state properties which define the infinite-frequency xc kernel have been found to be:
(i) Particle density; (ii) The kinetic stress tensor; (iii) The pair-correlation function; (iv)The exchange-correlation potential.
The former three can be found, in principle exactly, by the method of the quantum Monte-Carlo,
while the latter can be obtained by the inversion of the Kohn-Sham equations after the ground-state density has been found.

Keeping in view applications to finite and non-periodic infinite systems,
we have worked out the real-space form of the high frequency xc kernels.
The real space formulation is most directly connected with the quantum continuum mechanics formulation of Ref. \onlinecite{Tao-09}, and clearly displays the physical meaning of the anti-adiabatic $f_{xc}$ as elastic constants (stiffnesses) of the system. 

In the periodic case and in the long-wave limit, our results  contain an important singularity in the scalar $f_{xc}$ as a function of the wave-vector.
This singularity is known to be a manifestation of the ultra non-locality  of $f_{xc}$ in space and its presence is crucial for reproducing the excitonic effect in semiconductors and insulators by means of TDDFT.

We expect the results of this work to facilitate the construction of exchange-correlation kernels of TDDFT and TDCDFT, accurate enough to
account for the many-body effects in the linear response theory with applications to optics, electron energy-loss spectroscopy, and other fields.  
In particular, approximate formulas may be constructed for the frequency dependence of the xc kernels,  interpolating
between the adiabatic and the anti-adiabatic limits. In any case, it remains an interesting question to fully understand the relation between the adiabatic and the anti-adiabatic approximation to the xc kernels.  They are formally very similar, instantaneous functionals of displacement and density respectively, but the physics they embody is quite different since the system is assumed to be fully relaxed to the instantaneous ground-state in the former case, while, in the latter, the elastic deformation dictates the instantaneous state of the system.  

\section{Acknowledgements}
This work was supported by DOE grant DE-FG02-05ER46203 (GV, SP). 
GV gratefully acknowledges the hospitality of the ETSF in San Sebastian where this work was initiated and the Institute for Solid State Physics of Tokyo   where it was completed.
VUN gratefully acknowledges the hospitality of the University of Missouri-Columbia. IVT acknowledges support by the Spanish MEC (FIS2007-65702-C02-01), ACI-Promociona (ACI2009-1036) Grupos Consolidados UPV/EHU del Gobierno Vasco" (IT-319-07), the European Union through e-I3 ETSF project (Contract Number 211956).\\

\appendix

\section{Derivation of $M^{xc}_{\mu\nu}(\rv,\rv')$}
\label{D_Mxc}
In this Appendix, we outline the derivation of the expectation values of the commutators relevant to the evaluation of 
the moment by Eq.~(\ref{Main_comm}).  Since the system is assumed paramagnetic, the equilibrium expectation value of the current is set to  zero throughout.

(i) Kinetic energy part. A straightforward evaluation leads to 
\bern
&&\frac{m}{2} \langle \left[j_{n\mu}(\rv),\left[  p_n^2,j_{n\nu}(\rv') \right] \right] \rangle
=  \pa'_\alpha
\pa_\alpha  \delta(\rv-\rv') \langle p_{n\mu} j_{n\nu}(\rv) \rangle  
\cr\cr  
&+& \pa'_\alpha \pa_\nu  \delta(\rv-\rv')  \langle p_{n\alpha} j_{n\mu}(\rv)  \rangle -  \pa'_\alpha \pa'_\mu  \delta(\rv-\rv')  \langle p_{n\alpha} j_{n\nu}(\rv)  \rangle\nn\\ 
&+&\frac{1}{4 m}  \pa'_\alpha \pa_\nu \pa'_\mu  \pa_\alpha   \delta(\rv-\rv') \langle \delta(\rv-\rv_n)\rangle.
\eern
Introducing a tensor
\bern
T_{\mu\nu}(\rv)=\frac{1}{ 2 } \sum\limits_n \langle p_{n \mu} j_{n \nu}(\rv) +
p_{n \nu} j_{n \mu}(\rv) \rangle,
\eern
we express it as
\bern
T_{\mu\nu}(\rv)= 
- \frac{1}{m} \left.  \pa'_\mu \pa'_\nu  \rho_1(\rv,\rv') \right|_{\rv'=\rv} + \frac{1}{2 m} \pa_\mu \pa_\nu  n_0(\rv),
\eern
where $\rho_1(\rv,\rv')$ is the one particle density matrix. 
For the kinetic energy part we, therefore, have
\ber
&&m^2 \langle \left[j_{\mu}(\rv),\left[ T ,j_{\nu}(\rv') \right] \right] \rangle
=  
\pa_\alpha \pa'_\alpha \delta(\rv-\rv') T_{\mu\nu}(\rv)  + 
\cr\cr   &&\pa'_\alpha  \pa_\nu  \delta(\rv-\rv')  T_{\alpha\mu}(\rv)   - \pa'_\mu \pa'_\alpha  \delta(\rv-\rv')  T_{\alpha\nu}(\rv) \nn\\ 
&+&\frac{1}{4 m} \pa'_\alpha \pa_\nu \pa'_\mu  \pa_\alpha  \delta(\rv-\rv') n_0(\rv).
\label {T_e}
\eer

(ii) The external potential part is easily evaluated to
\ber
&&m^2 \left\langle \left[j_{\mu}(\rv),\left[ \sum\limits_n V_0(\rv_n) ,j_{\nu}(\rv') \right] \right] \right\rangle
= \nn\\&& n_0(\rv)  \pa_\mu \delta(\rv-\rv') \pa_\nu  V_0(\rv).
\label{V0_e}
\eer

(iii) For the interaction part, we need the two types of   commutators
\begin{widetext}
\bern
m^2 \left[j_{1\mu}(\rv),\left[V_C(|\rv_1-\rv_2|),j_{1\nu}(\rv') \right] \right]= \delta(\rv-\rv_1) \pa_\mu \delta(\rv-\rv') \pa'_\nu V_C(|\rv'-\rv_2|),\\
m^2 \left[j_{1\mu}(\rv),\left[V_C(|\rv_1-\rv_2|),j_{2\nu}(\rv') \right] \right]
= \delta(\rv-\rv_1) \delta(\rv'-\rv_2)  \pa_\mu \pa'_\nu V_C(|\rv-\rv'|).
\eern
Then the interaction part can
 be rewritten in terms of the pair correlation function $g(\rv,\rv')$ of Eq.~(\ref{g}) as
\ber
&&m^2 \left\langle \left[j_{\mu}(\rv),\left[ \frac{1}{2} \sum\limits_{n\ne m} V_C(|\rv_n-\rv_m|) ,j_{\nu}(\rv') \right] \right] \right\rangle
= n_0(\rv) \times \cr\cr  
&&\left[ \int  [g(\rv,\rv'')-1] \pa_\mu \delta(\rv-\rv') \pa'_\nu  n_0(\rv'') V_C(|\rv'-\rv''|) d\rv''
+   n_0(\rv') [g(\rv,\rv')-1]  \pa_\mu \pa'_\nu V_C(|\rv-\rv'|) \right.  \cr\cr
&&\left. +      \pa_\mu \delta(\rv-\rv') \pa'_\nu V_H(\rv')
+ n_0(\rv')  \pa_\mu \pa'_\nu V_C(|\rv-\rv'|)\right], \ \ \ \
\label{C_e}
\eer
\end{widetext}
where 
\be
V_H(\rv)=\int n_0(\rv') V_C(|\rv-\rv')|) d\rv' \nn
\ee
is Hartree potential.

(iv) A considerable simplification comes from the Newton's law 
\ber
&&\pa_\nu T_{\nu\mu}(\rv) +n_0(\rv)\pa_\mu V_0(\rv) \nn\\&&+ n_0(\rv) \int g(\rv,\rv') \pa_\mu \frac{e^2 n_0(\rv')}{|\rv-\rv'|} d\rv'' =0,
\label{Nl_e}
\eer
the latter immediately following from the obvious relation
$
\langle [j_\mu(\rv),H_0]\rangle=0
$.

(v) $M_{\mu\nu}(\rv,\rv')$ is obtained by combining Eqs.~(\ref{T_e}), (\ref{V0_e}), (\ref{C_e}), and (\ref{Nl_e}).
Then $M^s_{\mu\nu}(\rv,\rv')$ is constructed from  $M_{\mu\nu}(\rv,\rv')$ by replacing  $V_0$ with $V_S=V_0+V_H+V_{xc}$ 
and setting the interaction term to zero. We finally arrive at Eq.~(\ref{fxc_fin}) by the use of Eq.~(\ref{Mxc}).

\section{The anti-adiabatic limit of the xc kernel for the homogeneous electron gas}
\label{Stefano}
The tensor xc kernel of a homogeneous electron gas has the form\cite{Vignale-96} 
\begin{equation}\label{A1}
[\hat f_{xc}^{h}(q,\omega)]_{ij} = \frac{q^2}{\omega^2}\left[f_{xcL}^{h}(q,\omega)\hat q_i\hat q_j + f_{xcT}^{h}(q,\omega)  (\delta_{ij}-\hat q_i\hat q_j)\right]\,,
\end{equation}
where $\hat \qv$ is the unit vector in the direction of $\qv$.  

The longitudinal (L) component of the kernel coincides with the scalar xc kernel.  
In the high-frequency limit, 
by using the f-sum rule and the third-moment sum rule for the density-density response functions (or the
first-moment sum rule for the current-current response functions) for the interacting and non-interacting gas,\cite{Giuliani&Vignale} one readily obtains 
\begin{equation}\label{A2}
f_{xcL}^{h}(q,\infty) =  f^{(1)}_{xcL}(q,\infty) + f^{(2)}_{xcL}(q,\infty)
\end{equation}
where
\begin{equation}\label{A3}
f^{(1)}_{xcL}(q,\infty) = \frac{4}{3} \alpha r^2_s   
\int_{0}^{\infty}  dk ~  \left[ S(k,r_s) - 1 \right]    f\left( k/q \right)
\end{equation}
and
\begin{equation}\label{A4}
f^{(2)}_{xcL}(q,\infty) = \frac{8}{3} \pi r^3_s t_c(r_s)\;,
\end{equation}
with
\begin{equation}\label{A5}
f\left( x \right) = \frac{x^2}{2} \left[ \frac{5}{3} - x^2 + \frac{\left( x^2-1 \right)^2}{2 x} \ln\left( \Big{|} \frac{1 + 1/x}{1  - 1/x}  \Big{|} \right)\right]\;,
\end{equation}
$t_c(r_s)$ is the correlation kinetic energy, and $S(k,r_s)$ is the static structure factor. 
We emphasize  that the contribution due to the correlation kinetic energy, $f^{(2)}_{xcL}(q,\infty)$, is independent of $q$.
In the expressions above the atomic units  in which $\hbar = e^2 = m = 1$  
are used and the wave vectors are expressed in units of the Fermi wave vector $q_F = \alpha / r_s$,
with $\alpha = (9 \pi/4)^{1/3}$.

In the same limit, the transverse component is given by 
\begin{equation}\label{A6}
f_{xcT}^{h}(q,\infty) =   -\frac{1}{2}f^{(1)}_{xcL}(q,\infty) + \frac{1}{3}  f^{(2)}_{xcL}(q,\infty)
\end{equation}

\begin{figure}
\includegraphics[width=1.0\columnwidth]{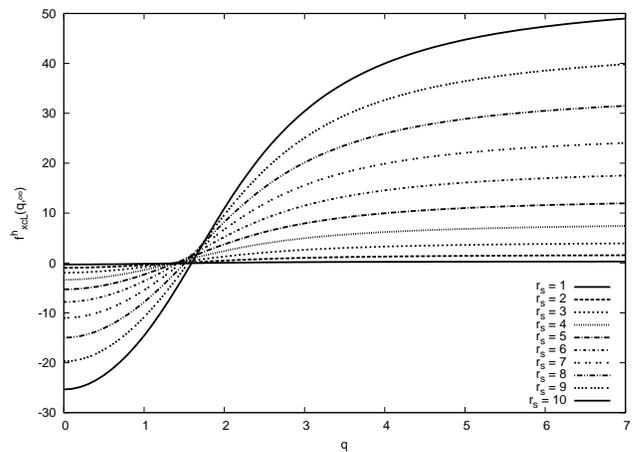}
\caption{High-frequency limit of the longitudinal xc-kernel for the homogenous electron gas plotted as function of  $q$ for different values of $r_s$.
Asymptotically the only finite contribution is due to the term containing the correlation kinetic energy.}
\label{fig1}
\end{figure}

\begin{figure}
\includegraphics[width=1.0\columnwidth]{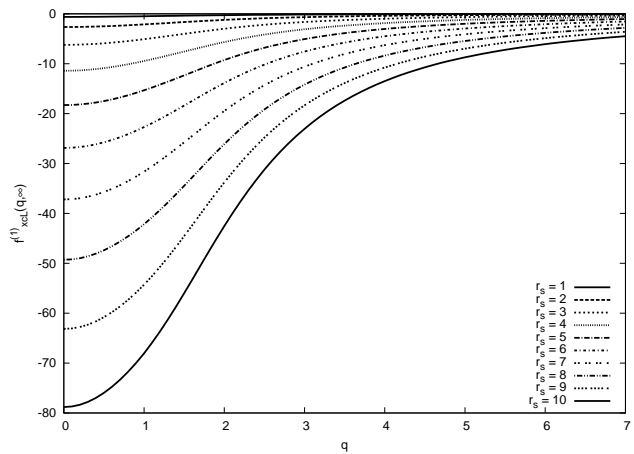}
\caption{Here the quantity in Fig. (\ref{fig1}) is plotted by neglecting  $f^{(2)}_{xcL}(q,\infty)$ [Eq. (\ref{A4})]. }
\label{fig2}
\end{figure}

\begin{figure}
\includegraphics[width=1.0\columnwidth]{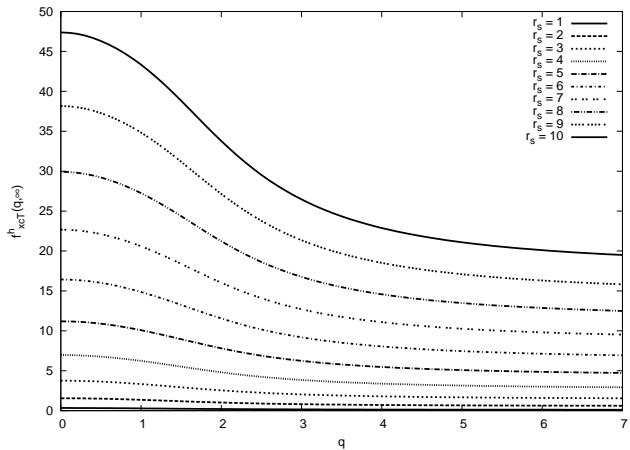}
\caption{High-frequency limit of the transverse xc-kernel for the homogenous electron gas plotted as function of  $q$ for different values of $r_s$.
Asymptotically the only finite contribution is due to the term containing the correlation kinetic energy.}
\label{fig3}
\end{figure}

For the calculation of $S(k,r_s)$ we have employed the analytic static structure factor provided in Ref.~\onlinecite{Gori-Giorgi-00},
and the correlation kinetic energy has been calculated from the parametrized correlation energy of Appendix B
of the same reference, by making use of the virial theorem. The resulting quantities $f_{xcL}^{h}(q,\infty)$, $f_{xcT}^{h}(q,\infty)$ and $ f^{(1)}_{xc}(q,\infty) $
are plotted in Fig.~(\ref{fig1}), Fig.~(\ref{fig2}) and Fig.~(\ref{fig3})  respectively.  Among other facts, these plots show that 
the contribution due to the correlation kinetic energy  is very important at any $q$ -- a fact already pointed out by Iwamoto and Gross\cite{Iwamoto87} for  the longitudinal component at $q=0$.   Moreover, it is clear that $ f^{(2)}_{xcL}(q,\infty)$ can be identified as the $q \to \infty$ limit of $f^{h}_{xcL}(q,\infty)$. 
Overall, $f_{xcL(T)}^{h}(q,\infty)$ has a strong dependence on both $q$ and $r_s$.  
From Fig.~(\ref{fig1}) and Fig.~(\ref{fig3}), it is evident that $f^{h}_{xcL}(q,\infty)$ and
$f^{h}_{xcT}(q,\infty)$ have very different ranges of values. 
We observe that while
$f^{h}_{xcT}(q,\infty)$ is always
positive, $f_{xcL}^{h}(q,\infty)$ changes sign at about a fixed value of $q/k_F$ for all  $r_s$:  this suggests
that  the $r_s$-dependence of $f_{xcL}^{h}(q,\infty)$, with $q$ expressed in units of $k_F$ may be an overall scale factor.
Clearly, the fact that $f^{h}_{xcL(T)}(q,\infty)$ go to a constant in the large $q$ limit implies that the corresponding local field factors diverge as $q^2$.
This is also observed at finite frequency, and, indeed all the way to zero frequency, as  first noticed by Holas\cite{Holas86} 
and then confirmed by quantum Monte Carlo\cite{Moroni95} and diagrammatic calculations.\cite{Vignale88}  In particular, the exact relation 
$f^{h}_{xcL}(q \to \infty ,\infty) = -3 f^{h}_{xcL}(q \to \infty ,0)$~\cite{Holas86} is well satisfied.

\end{document}